\documentclass[twocolumn]{aastex631}
\usepackage{graphicx}	
\usepackage{amsmath}	
\usepackage{amssymb}	

\received{XXX}
\revised{YYY}
\accepted{ZZZ}

\submitjournal{ApJ}

\shorttitle{IC~4296: Threads, Ribbons, and Rings}
\shortauthors{Condon et al.}

\begin{document}

\title{Threads, Ribbons, and Rings in the Radio Galaxy IC~4296}

\correspondingauthor{J.~J.~Condon}
\email{jcondon@nrao.edu}

\author[0000-0003-4724-1939]{J.~J.~Condon}
\affiliation{National Radio Astronomy Observatory,
             520 Edgemont Road,
             Charlottesville, VA 22903, USA}

\author[0000-0001-7363-6489]{W.~D.~Cotton}
\affiliation{National Radio Astronomy Observatory,
             520 Edgemont Road,
             Charlottesville, VA 22903, USA}
\affiliation{South African Radio Astronomy Observatory,
             2 Fir Street, Black River Park,
             Observatory 7925, South Africa}

\author[0000-0002-2340-8303]{S.~V.~White}
\affiliation{Department of Physics and Electronics, Rhodes University,
  PO Box 94, Grahamstown 6140, South Africa}

\author[0000-0001-5205-8501]{S.~Legodi}
\affiliation{South African Radio Astronomy Observatory,
             2 Fir Street, Black River Park,
             Observatory 7925, South Africa}

\author[0000-0003-3636-8731]{S.~Goedhart}
\affiliation{South African Radio Astronomy Observatory,
             2 Fir Street, Black River Park,
             Observatory 7925, South Africa}

\author{K.~McAlpine}
\affiliation{South African Radio Astronomy Observatory,
             2 Fir Street, Black River Park,
             Observatory 7925, South Africa}

\author{S.~M.~Ratcliffe}
\affiliation{South African Radio Astronomy Observatory,
             2 Fir Street, Black River Park,
             Observatory 7925, South Africa}
\affiliation{Tsolo Storage Systems,
             12 Links Drive,
             Pinelands 7405, South Africa}

\author[0000-0002-1873-3718]{F.~Camilo}
\affiliation{South African Radio Astronomy Observatory,
             2 Fir Street, Black River Park,
             Observatory 7925, South Africa}

\begin{abstract}
  The nearby elliptical galaxy IC~4296 has produced a large (510\,kpc)
  low-luminosity radio source with typical FR\,\textsc{i}
  core/jet/lobe morphology.  The unprecedented combination of
  brightness sensitivity, dynamic range, and angular resolution of a
  new 1.28\,GHz MeerKAT continuum image reveals striking new
  morphological features which we call threads, ribbons, and rings.
  The threads are faint narrow emission features originating where
  helical Kelvin-Helmholtz instabilities disrupt the main radio jets.
  The ribbons are smooth regions between the jets and the lobes,
  and they appear to be relics of jets powered by earlier activity
  that have since come into pressure equilibrium.  Vortex rings in the
  outer portions of the lobes and their backflows indicate that the
  straight outer jets and ribbons are inclined by $i =60^\circ \pm
  5^\circ$ from the line-of-sight, in agreement with photometric,
  geometric, and gas-dynamical estimates of inclination angles near
  the nucleus.
\end{abstract}

\keywords{Elliptical Galaxies (456) -- Interstellar magnetic fields
  (845) -- Radio galaxies (1343) -- Radio jets (1347)}



\section{Introduction}

\object{IC 4296} is the brightest of 18 galaxies comprising the Nearby
Optical Galaxy Group 722 in the cluster Abell 3565 \citep{giu00}.  It
is a purely elliptical galaxy with a S\'ersic brightness profile and
no detectable exponential disk related to star formation
\citep{don11}.  Its infrared color
\begin{equation}
  \log\biggl( \frac {L_{24\,\mu\mathrm{m}}} {\mathrm{erg\,s}^{-1}}\biggr) -
  \log \biggl( \frac {L_{K\mathrm{s}}} {L_{K\mathrm{s},\odot}}\biggr)  = 30.16
\end{equation}
between $\lambda = 24\,\mu\mathrm{m}$ and $\lambda =
2.2\,\mu\mathrm{m}$ is typical of ``red and dead'' gas-free galaxies
with no evidence of recent star formation, and most of its
$24\,\mu\mathrm{m}$ luminosity is likely circumstellar dust emission
from old giant stars \citep{tem09}.  Even if all of its far-infrared
flux density $S_{70\,\mu\mathrm{m}} = 118 \pm 12\,\mathrm{mJy}$
\citep{tem09} were attributed to star formation consistent with the
far-infrared/radio correlation, the 1.4\,GHz flux density produced by
star formation in IC~4296 would be a negligible $S < 1$\,mJy.

IC~4296 is at heliocentric redshift $z = 0.01247 \pm 0.00003$
\citep{smi00}. However, its J2000 position $\alpha =
13^\mathrm{h}\,36^\mathrm{m}\, 39\,\fs045$, $\delta = -33^\circ \,
57\arcmin\, 56\,\farcs91$ \citep{skr06} is near the anticenter of the
supergalactic plane ($\mathrm{SGL} = 151\fdg8$, $\mathrm{SGB} =
-0\,\fdg4$) where the redshift distance may differ significantly from
the actual distance.  Velocity-independent angular-size distances have
been estimated from surface-brightness fluctuations of IC~4296 by
\citet{lau98} ($D_\mathrm{A} = 49.4 \,\mathrm{Mpc}$) and by
\citet{mei00} ($D_\mathrm{A} = 49\,\mathrm{Mpc}$).  For simplicity we
adopt the comoving distance $D_\mathrm{C} = D_\mathrm{A} (1 + z) =
50\,\mathrm{Mpc}$ so $1\arcsec = 240\,\mathrm{pc}$ and $1\arcmin =
14.4\,\mathrm{kpc}$.

\citet{mil60} first identified IC~4296 with an extended (largest
angular size LAS $>50\arcsec$) radio source.  The radio emission
powered by IC~4296 is actually so extended (LAS $ \sim 36\arcmin$)
that it comprises three sources in the Parkes 2.7\,GHz catalog
\citep{wri90}: B1332$-$336 (northwest lobe), B1333$-$336 (core and
inner jets), and B1334$-$338 (southeast lobe).  Multifrequency radio
images with various angular resolutions show that the radio source
consists of an unresolved core on the nucleus of IC~4296, symmetric
curved jets, and slightly edge-brightened lobes \citep{kil86,gro19}.

Our new MeerKAT 1.28\,GHz continuum images are the first having the
combination of high surface-brightness sensitivity and angular
resolution needed to display striking morphological features that we
call threads, ribbons, and rings.  This paper describes the MeerKAT
observations and data reduction (Section~\ref{sec:ObsAnalysis}),
presents the new total-intensity, spectral index, and polarization
images (Section~\ref{sec:Results}), and analyzes the new morphological
features (Section~\ref{sec:Discussion}).  Section~\ref{sec:Summary}
summarizes these results and discusses their wider significance for radio
astronomy and future radio telescopes.

Absolute quantities were calculated for a $\Lambda$CDM universe with
$H_0 = 70 \mathrm{~km~s}^{-1} \mathrm{~Mpc}^{-1}$ and
$\Omega_\mathrm{m} = 0.3$ from equations in \citet{con18}.  We use
the spectral-index sign convention $\alpha \equiv + d \log(S) /
d\log(\nu)$.

\section{Observations and Data Analysis \label{sec:ObsAnalysis}}

The 64 antenna MeerKAT array \citep{jon16,cam18,mau20} observed
IC~4296 during two sessions: 2020 March 16
from 18:47 UTC to 2020 March 17 04:50 UTC (60 antennas) and 2020 May
13 from 15:32 UTC to 2020 May 14 1:30 UTC (61 antennas), for a total
of 14.9 hours on target.  The sources PKS~0408$-$65 = J0408$-$6545 and
PKS~B1934$-$638 were used as the flux density, bandpass, and delay
calibrators; 3C~138 and 3C~286 were the polarization calibrators; and
PKS~1320$-$446 = J1323$-$4452 was the astrometric calibrator.  The
observing sequence cycled between J1323$-$4452 (1 minute) and IC~4296
(15 minutes) plus a flux/bandpass calibrator (10 minutes) every 2
hours.  Our flux-density scale is based on the \citet{rey94} spectrum
of PKS~B1934$-$638:
\begin{eqnarray}
  \log(S) = & -30.7667 + 26.4908 (\log\nu)
  - 7.0977 (\log \nu)^2 \nonumber \\
   & +0.605334 (\log\nu)^3\, , \qquad\qquad\qquad\qquad\qquad
\end{eqnarray}
where $S$ is the flux density in Jy and $\nu$ is the frequency in MHz.
We used 8\,s integration times divided into 4096 spectral channels
between 856 and 1712\,MHz to minimize time and bandwidth smearing.
The useful frequency range limited by radio-frequency response and
filters is $\approx 880$ to 1670\,MHz.  Data flagging and calibration
were performed as described in \citet{mau20} and \citet{cot20}, with
each session calibrated independently.

\subsection{Imaging}

The wide-band, wide-field imager MFImage in the \emph{Obit}
package\footnote{https://www.cv.nrao.edu/$\sim$bcotton/Obit.html}
\citep{obi08} was used as described in \citet{mau20} and \citet{cot20}
except as noted below.  MFImage \citep{cot18} divides the imaged
region into small facets to approximate the non-coplanarity
\citep{per99} of the sky, and multiple frequency bins are imaged
independently and CLEANed jointly to accommodate smooth variations of
the sky and the antenna pattern with frequency.

MeerKAT's shortest baselines ($\geq 29$\,m) double in wavelengths
between 856 and 1712\,MHz, potentially leading to a variable fraction
of the total intensity from the very large radio source being
recovered as a function of frequency, a frequency-dependent negative
``bowl'' in the radio image, and an artificially steepened source
spectrum.  To minimize these effects, we applied a Gaussian taper with
a root-mean-square (rms) width $\sigma = 500$ wavelengths to the
projected baselines at each frequency.

\begin{figure*}[!htb]
\centerline{
  \includegraphics[width=\textwidth,trim = {1.2cm 6.2cm 1.6cm 6.cm},
    clip]{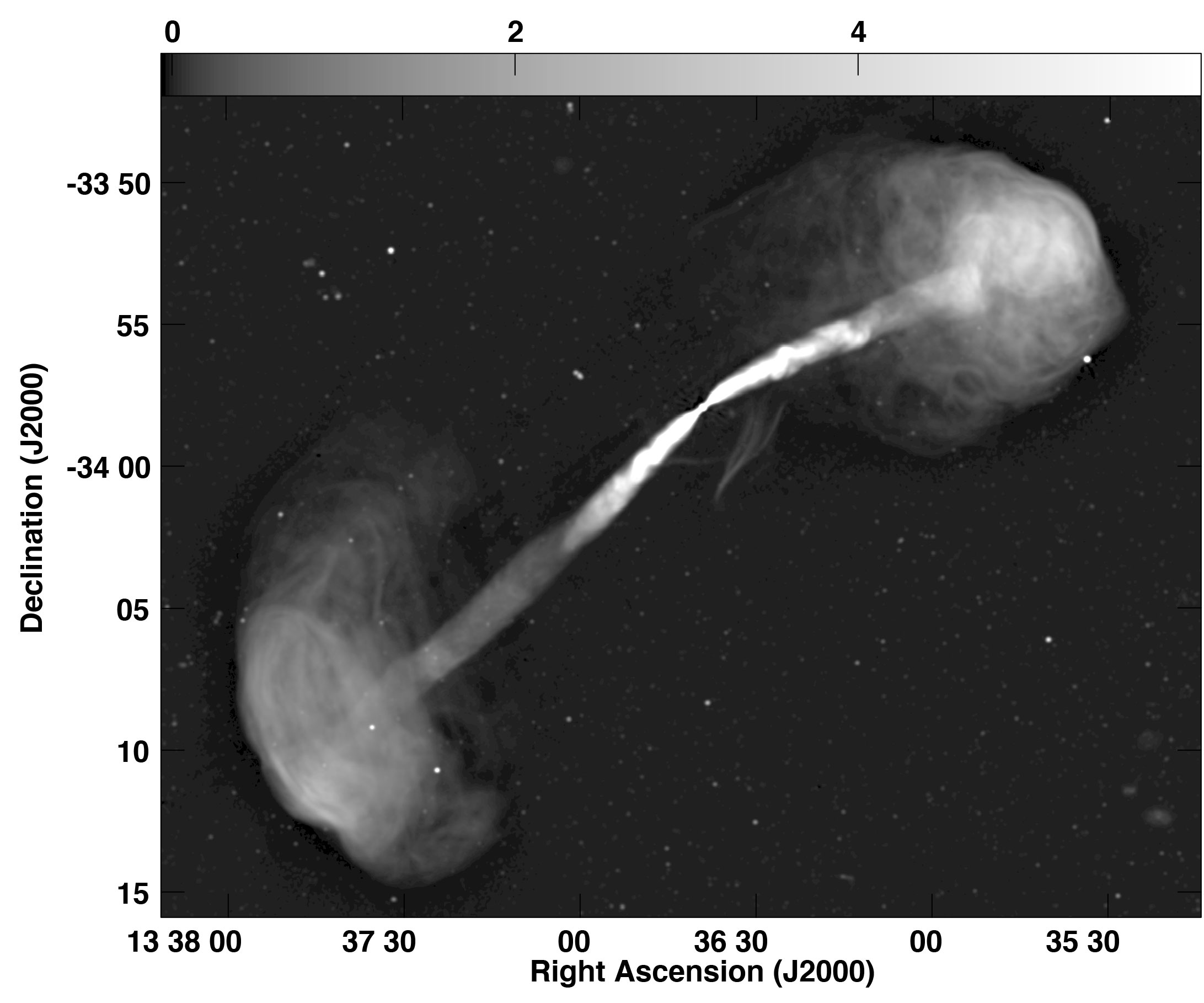}
}
\caption{This gray-scale display shows the final total-intensity image of
  IC~4296 in the brightness range $-0.06 < S_\mathrm{p}\,
  (\mathrm{mJy\,beam}^{-1}) < +6$ as indicated by the linear scale bar
  at the top.
   The darkest streaks parallel to the southeast jet in this
   figure are display compression artifacts.
The brightness
distribution of the saturated inner jets is better seen in the left
panel of Figure~\ref{fig:JetsThreads}, and the faint ``threads''
emerging from the bright jets are emphasized in the right panel of
Figure~\ref{fig:JetsThreads}. } 
\label{fig:IC4296IPolAll}
\end{figure*}

\subsection{Deconvolution of Stokes Q and U}

Independently imaging Stokes Q and U is operationally simpler than a
joint deconvolution but loses some of the information available to a
combined deconvolution.  Averaging Q or U in a source with significant
Faraday rotation reduces the wide-band average polarization.  When
applied to IC~4296, independent deconvolution of Stokes Q and U also
destabilized the solution and produced large-scale artifacts.
Consequently we derived the polarized intensity
\begin{equation}
  P\ =\ \sqrt{Q^2\ +\ U^2}\, 
\end{equation}
in each frequency bin from a combination of the Q and U images and
used it to drive the deconvolution.  The polarized intensity should
vary smoothly with frequency and is relatively insensitive to Faraday
rotation if the data and imaging processes have adequate spectral
resolution.  The rms-weighted frequency average of
$P$ can be used to drive deconvolution by CLEAN.  The process used for
each CLEAN major cycle is:
\begin{enumerate}
\item {\bf Form dirty/residual images.} Dirty/residual images are
  formed in each of Q and U in each frequency bin and in each facet
  imaged.
\item {\bf Form polarized intensity.}  In each frequency bin and facet
  imaged, the Q and U images are combined into a polarized intensity
  plane.
\item {\bf Average polarized intensity.}  In each facet being imaged,
  a weighted average of the polarized intensity planes is made and
  used to drive CLEAN in both Q and U.  The weighting is proportional to
  $1/\mathrm{rms}$ in the Q and U frequency bin images.
\item {\bf Select CLEAN components.}  Clean components are selected by
  a \citet{cla80} inner CLEAN from the polarized intensity plane.
  Selected pixels are collected into lists with a limited portion of
  the dirty beam for each facet, frequency bin, and polarization.  At
  cells picked, the pixel values in the frequency-bin images for the Q
  and U residuals are recorded and the loop gain times the dirty beam
  is subtracted.  This is repeated until a stopping criterion is
  satisfied.
\item {\bf Subtract from the $u,v$ data.}  The collected sets of CLEAN
  components in Q and U are Fourier transformed and subtracted from
  the Q and U residual data sets.
\end{enumerate}

The polarization images were restored with elliptical Gaussian beams
having full width between half-maximum (FWHM) diameter $7\,\farcs16
\times 7\,\farcs01$ and major-axis PA $= 89\fdg41$ measured counterclockwise
from north.

\section{Imaging Results\label{sec:Results}}

\begin{figure*}[!htb]
\centerline{
  \includegraphics[width=0.57\textwidth,trim = {1.5cm 6.0cm 1.5cm 6.8cm},
    clip]{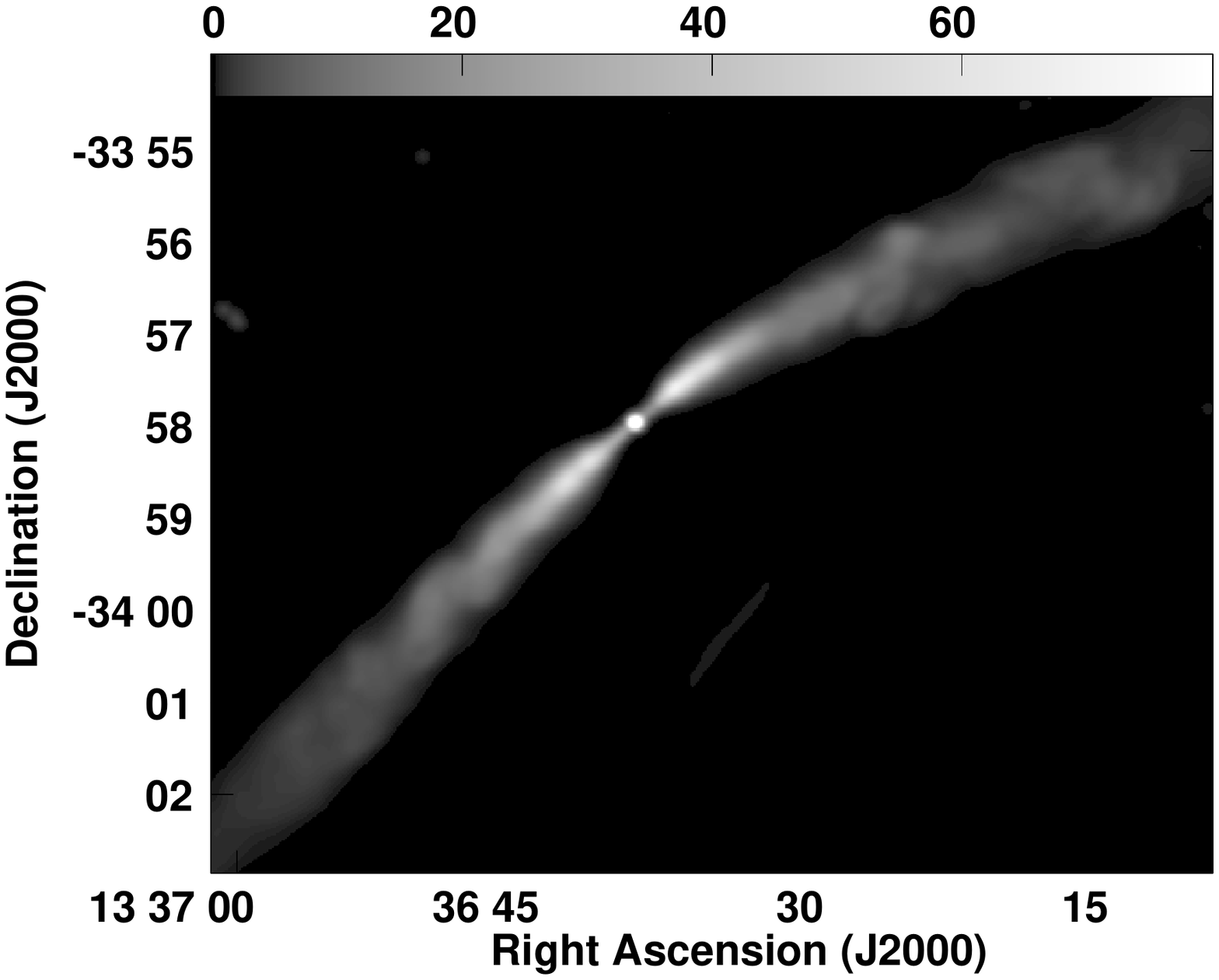}
  \includegraphics[width=0.42\textwidth,trim={.7cm 2.8cm 1.8cm 1.3cm},
    clip]{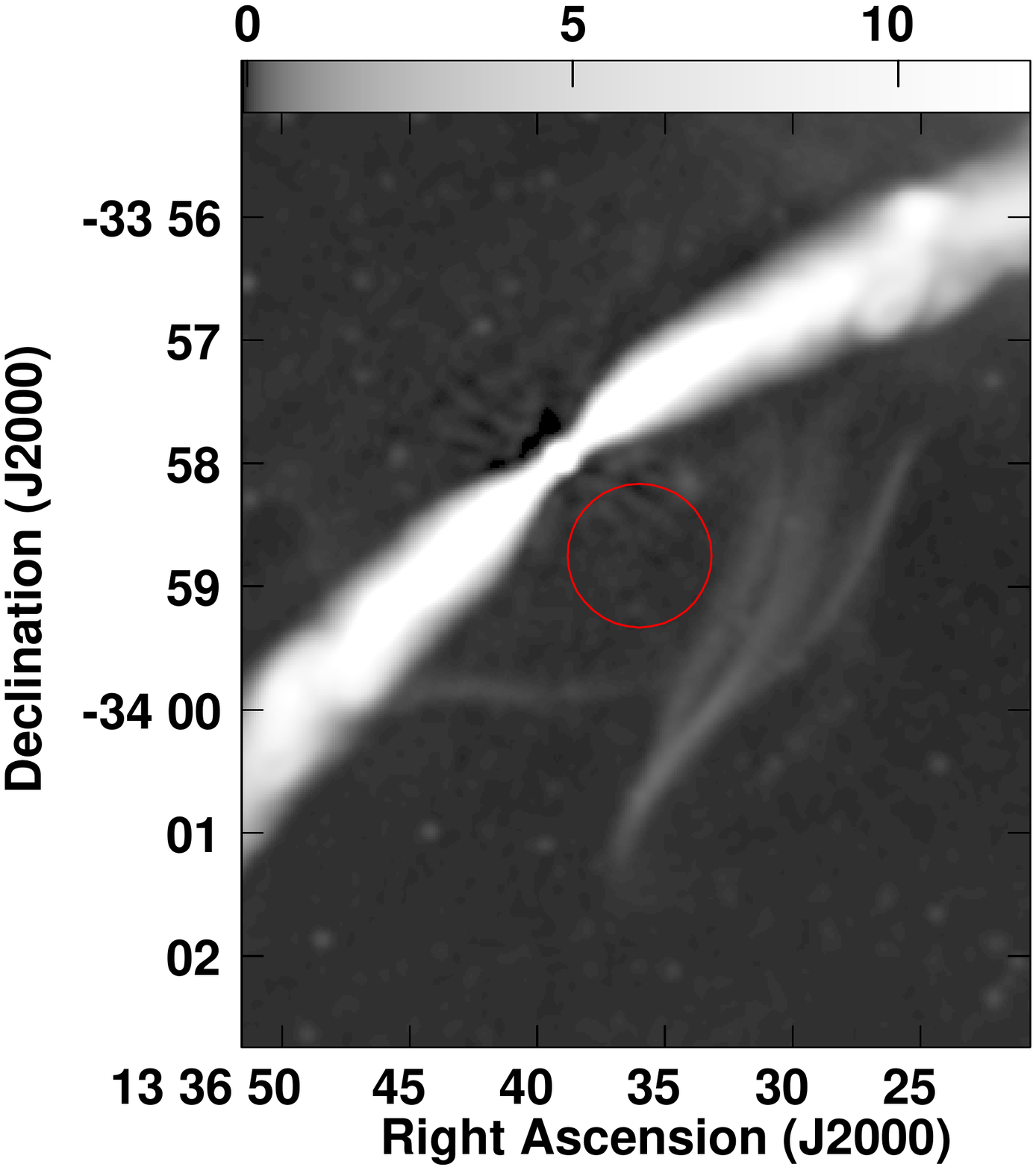}
}
\caption{The left panel shows the total-intensity image of the IC~4296
  core and jets with an expanded linear stretch between $-0.08$ and
  $+80\,\mathrm{mJy\,beam}^{-1}$.  The right panel brings
  out the faint threads with a logarithmic stretch between $-0.06$ and
  $+12\, \mathrm{mJy\,beam}^{-1}$.
    The dark lines radiating
    from the nucleus are faint  ($\sim -40\,\mu\mathrm{Jy\,beam}^{-1}$)
    imaging artifacts caused by residual phase errors.
The red circle with diameter
$70\arcsec$ and centered on $\alpha =
13^\mathrm{h}\,36^\mathrm{m}\,36^\mathrm{s}$, $\delta =
-33^\circ\,58\arcmin\,45\arcsec$ bounds the bright X-ray emission
shown by the red circle in fig.~4 of \citet{gro19}.}
\label{fig:JetsThreads}
\end{figure*}

The $(u,v)$ data were re-weighted in proportion to $1/\mathrm{rms}$
in the observed visibilities in 10 minute intervals, and the data from
the two observing sessions were combined.  The total-intensity data
were imaged over a circle of radius $1\fdg 2$ plus outlier facets up
to $1\fdg5$ from the pointing center to cover sources with attenuated
flux densities $S > 3 \,\mathrm{mJy}$ from the SUMSS 843\,MHz source
catalog \citep{mau03}. The total bandpass was divided into 14 subbands
with 5\% fractional bandwidths.  CLEANing with loop gain 0.03 was
stopped after $6.8 \times 10^6$
CLEAN components, leaving a maximum residual $S_\mathrm{p} =
71\,\mu\mathrm{Jy\,beam}^{-1}$.  The total CLEANed flux density at the
1.28\,GHz band center is 15.1\,Jy.  The 1.28\,GHz total-intensity
image was restored with an elliptical Gaussian beam having FWHM size
$6\,\farcs96 \times 6\,\farcs67$ and major-axis position angle PA =
$-88\,\fdg5$.

Rayleigh-Jeans brightness temperatures in this CLEAN image are related
to peak flux densities via $T_\mathrm{b}(\mathrm{K}) \approx
16.1\,S_\mathrm{p} (\mathrm{mJy\,beam}^{-1})$.  The rms fluctuation in
source-free areas near the center of the CLEAN image is $\sigma = 5.4
\,\mu\mathrm{Jy\,beam}^{-1} = 87\,\mathrm{mK}$.  The rms noise
measured near the first zero of the primary beam is only
$\sigma_\mathrm{n} = 2.0 \,\mu\mathrm{Jy\,beam}^{-1}$.  The ``rms''
confusion $\sigma_\mathrm{c}$ can be defined as half the range of peak
flux densities containing 68\% of the pixels in a noiseless image.  It
is $\sigma_\mathrm{c} = 1.8\pm 0.1 \,\mu\mathrm{Jy\,beam}^{-1}$ at
$\nu = 1.266\,\mathrm{GHz}$
in a $7\,\farcs 6$ FWHM circular Gaussian beam \citep{mat21}.  For
small changes in beam size it scales as beam solid angle, so we expect
$\sigma_\mathrm{c} \approx 1.5 \,\mu \mathrm{Jy\,beam}^{-1}$
in our $6\,\farcs96 \times 6\,\farcs67$ beam.  Thus the sensitivity of
our total-intensity image is limited  by dynamic
range.

\subsection{Astrometry}

Source positions in  the MeerKAT image were
compared with the \emph{Gaia} catalog available from the NASA/IPAC
Infrared Science Archive (IRSA).  There are 43 radio sources $<1^\circ$ from
the pointing center with attenuated flux densities $> 200\,\mu$Jy and
\emph{Gaia} matches offset by $<1\farcs5$. The rms scatter of the MeerKAT
minus \emph{Gaia} offsets is only $0\,\farcs13$ in both right
ascension and declination, but the mean offsets are $\Delta \alpha =
+0\,\farcs46 \pm 0\,\farcs02$, $\Delta \delta = -0\,\farcs09 \pm
0\,\farcs02$.  These unexpectedly large offsets may reflect errors in
the MeerKAT on-line coordinate calculations.  All MeerKAT positions
quoted in this paper have been corrected to the \emph{Gaia} frame.

\subsection{The Total-Intensity On-Sky Image}

The image was divided by the primary-beam attenuation pattern
\citep{mau20} to give on-sky total intensities.  The depth of the
negative bowl remaining after incomplete CLEANing was measured to be
$-2 \pm 1 \,\mu\mathrm{Jy\,beam}^{-1}$ by averaging over pixels near,
but not in, extended emission from IC~4296.  The bowl was removed by
adding $+2\,\mu\mathrm{Jy\,beam}^{-1}$ to all pixels to create the
final total-intensity image shown in Figure~\ref{fig:IC4296IPolAll}.
The bright unresolved core and the inner jets are saturated in
Figure~\ref{fig:IC4296IPolAll}, so they have been replotted with a
wider linear stretch in the left panel of
Figure~\ref{fig:JetsThreads}. The faint ``threads'' leading from both
the  jets are emphasized by a
logarithmic stretch in the right panel of
Figure~\ref{fig:JetsThreads}.

The outer edge of the northwest lobe is $15 \farcm 3 \approx 220\,
\mathrm{kpc}$ from the core, and the southeast lobe reaches $20 \farcm
3 \approx 290\,\mathrm{kpc}$ from the core.  The total LAS~$=35\,
\farcm 5$ corresponds to a 510\,kpc length projected onto the sky.

We estimated the integrated 1.28\,GHz flux density of IC~4296 by
summing brightnesses over the whole source using the \emph{AIPS} verb
TVSTAT; it is $S = 14.8 \pm 0.5$\,Jy over 19660 beam solid angles.  A
systematic zero-level error of $1\,\mu\mathrm{Jy\,beam}^{-1}$ would
produce only a 0.02\,Jy error in $S$, and the random noise error is
negligible.  We quadratically added a 3\% intensity-proportional error
to yield the total flux-density uncertainty $\sigma_S \approx
0.5$\,Jy.

\subsection{Spectrum}

\begin{figure}[!htb]
\centerline{
  \includegraphics[width=0.57\textwidth,trim = {0.cm 9.5cm 5.cm 8.5cm},
    clip]{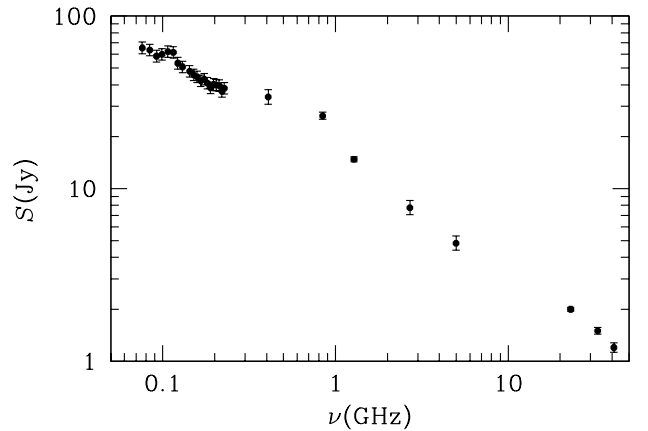}
}
\caption{ The radio spectrum of IC~4296 based on the flux densities
  listed in Table~\ref{tab:fluxes}. Abscissa: frequency (GHz).
  Ordinate: flux density (Jy) }
\label{fig:spectrum}
\end{figure}

The radio emission from IC~4296 is so extended that many of its
published flux densities are too low.  Even the selected flux
densities listed in Table~\ref{tab:fluxes} and plotted in
Figure~\ref{fig:spectrum} may be in error by more than their quoted
uncertainties.  The overall spectral index $\alpha = -0.70 \pm 0.08$
of IC~4296 is typical for an extended radio galaxy.

\begin{deluxetable}{r c l}
  \caption{IC~4296 flux densities}
  \label{tab:fluxes}
  \tablehead{
    $\nu$~~~~ & $S$ & Reference \\
    (GHz) & (Jy) & 
  }
  \startdata
  0.076 & $65.4 \pm 5.2$ & \citep{whi20} G4Jy 1080\\
  0.084 & $63.6 \pm 5.1$ & \\
  0.092 & $58.6 \pm 4.7$ & \\
  0.099 & $60.1 \pm 4.8$ & \\
  0.107 & $62.3 \pm 5.0$ & \\
  0.115 & $61.5 \pm 4.9$ & \\
  0.122 & $53.3 \pm 4.3$ & \\
  0.130 & $50.8 \pm 4.1$ & \\
  0.143 & $47.9 \pm 3.8$ & \\
  0.151 & $45.6 \pm 3.6$ & \\
  0.158 & $44.5 \pm 3.6$ & \\
  0.166 & $42.1 \pm 3.4$ & \\
  0.174 & $43.0 \pm 3.4$ & \\
  0.181 & $40.9 \pm 3.3$ & \\
  0.189 & $38.4 \pm 3.1$ & \\
  0.197 & $40.1 \pm 3.2$ & \\
  0.204 & $39.8 \pm 3.2$ & \\
  0.212 & $39.6 \pm 3.2$ & \\
  0.220 & $36.6 \pm 2.9$ & \\
  0.227 & $38.2 \pm 3.1$ & \\
  0.408 & $34.0 \pm 3.4$ & \citep{wri90} \\
  0.843 & $26.5 \pm 1.3$ & \citep{all14} \\
  1.28  & $14.8 \pm 0.5$ & (this paper) \\
  2.7   & $7.76 \pm 0.7\rlap{8}$ & \citep{wri90} \\
  5.0   & $4.83 \pm 0.4\rlap{8}$ & \citep{wri90} \\
  23    & ~~~$2.0 \pm 0.05$ & \citep{gro19} \\
  33    & ~~~$1.5 \pm 0.07$ & \citep{gro19} \\
  41    & ~~~$1.2 \pm 0.08$ & \citep{gro19} \\
  \enddata
\end{deluxetable}

\begin{figure*}[!htb]
\centerline{
  \includegraphics[width=\textwidth,trim = {0.cm 0.cm 0.cm 0.cm},
    clip]{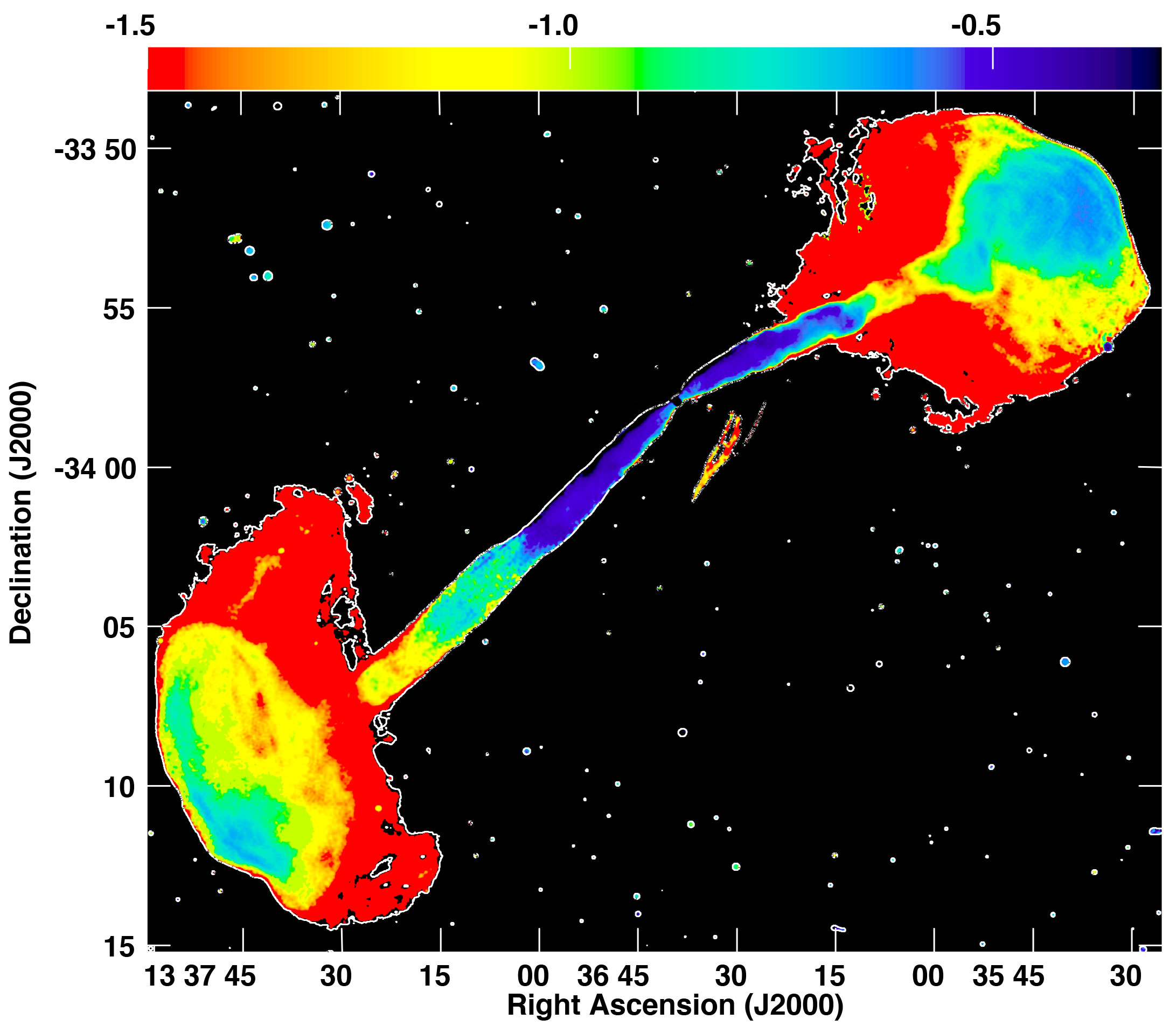}
}
\caption{This false color (scale bar at the top) spectral-index image
  of IC~4296 includes a single white total-intensity contour at
  $S_\mathrm{p} = 100\,\mu\mathrm{Jy\,beam}^{-1}$. The unresolved core
  (black) has $\alpha > 0$, the main jets have $\alpha \approx -0.5$,
  the ``ribbons'' have steeper spectra $\alpha \approx -0.8$, the
  outer edges of the lobes have ``normal'' spectra $\alpha\approx
  -0.7$, and the lobe backflows have the steepest spectra $\alpha <
  -1$.  }
\label{fig:IC4296SI}
\end{figure*}

The MeerKAT L-band spectral-index distribution within IC~4296 is
displayed as a false-color image in Figure~\ref{fig:IC4296SI}.  The
spectrum steepens going from the jets to the ribbons, lobes, and the
backflow regions behind the lobes.  The gradient in brightness is high
and the spectrum flattens in the brighter compression regions
just inside the outer edges of the lobes.  The very steep ($\alpha < -1$)
spectra of the backflow regions is expected from old electrons
experiencing synchrotron and inverse-Compton radiation losses.

\subsection{Polarization}

\begin{figure*}[!htb]
\centerline{
  \includegraphics[width=\textwidth,trim = {0.cm 0.cm 0.cm 0.cm},
    clip]{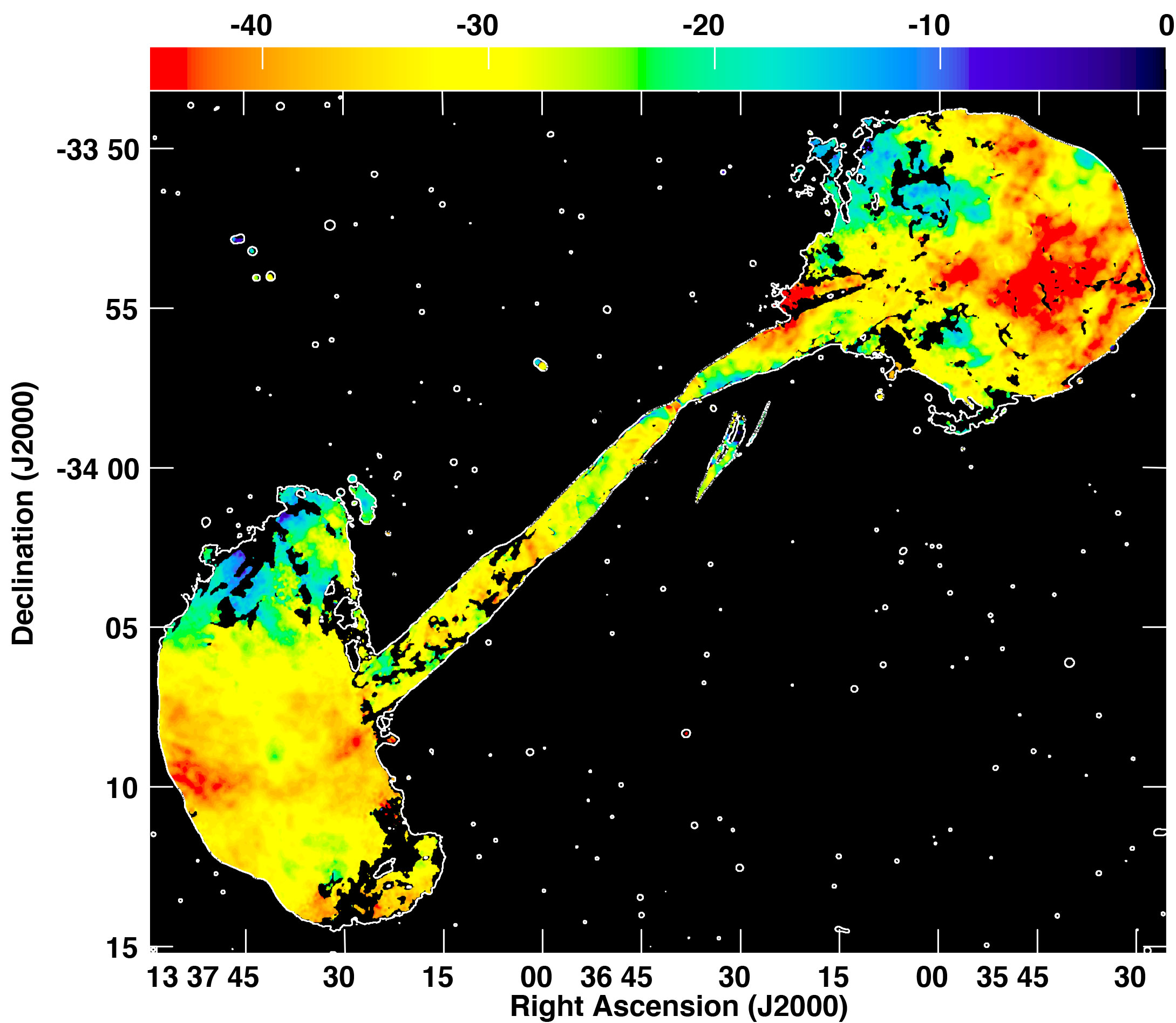}
}
\caption{False-color RM image of IC~4296 with the scale bar in
  rad\,m$^{-2}$ shown at the top.  One white total-intensity contour
  is plotted at $S_\mathrm{p} = 100 \,\mu
  \mathrm{Jy\,beam}^{-1}$. Most of this Faraday rotation is probably
  produced by the Galactic foreground, and not by any magnetized medium
  surrounding IC~4296.}
\label{fig:IC4296RM}
\end{figure*}

A rotation-measure (RM) fit was performed in each pixel by doing a
direct search over RM.  The test Faraday rotation that gives the
highest averaged, unwrapped polarized intensity was taken to be the
Faraday rotation at that pixel, the unwrapped polarization angle
extrapolated to zero wavelength was taken to be the intrinsic
polarization angle, and the maximum polarized intensity taken to be
the polarized intensity in that pixel.  This is essentially using the
peak of the Faraday synthesis \citep{bre05}.  The RMs
across IC~4296 are shown in the false-color Figure~\ref{fig:IC4296RM}.
The RMs in the bulk of the jets and lobes range between $-25$ and
$-45\,\mathrm{rad\,m}^{-2}$, with the northern portions of the lobes
around $-15\,\mathrm{rad\,m}^{-2}$.  The thread RMs range from $-30$
to $+8\,\mathrm{rad\,m}^{-2}$.  The mean Galactic foreground obtained
from the \citet{tay09} RM catalog of NVSS \citep{con98} sources is
$\langle \mathrm{RM} \rangle = -35 \pm 4 \,\mathrm{rad\,m}^{-2}$
within $2^\circ$ of IC~4296 at $l = 313\,\fdg5, \, b = +28\,\fdg0$,
so it appears to account for most of the observed RM.

\begin{figure}[!htb]
\centerline{
  \includegraphics[width=0.5\textwidth,trim = {1.cm 3.cm 0.4cm 3.cm},
    clip]{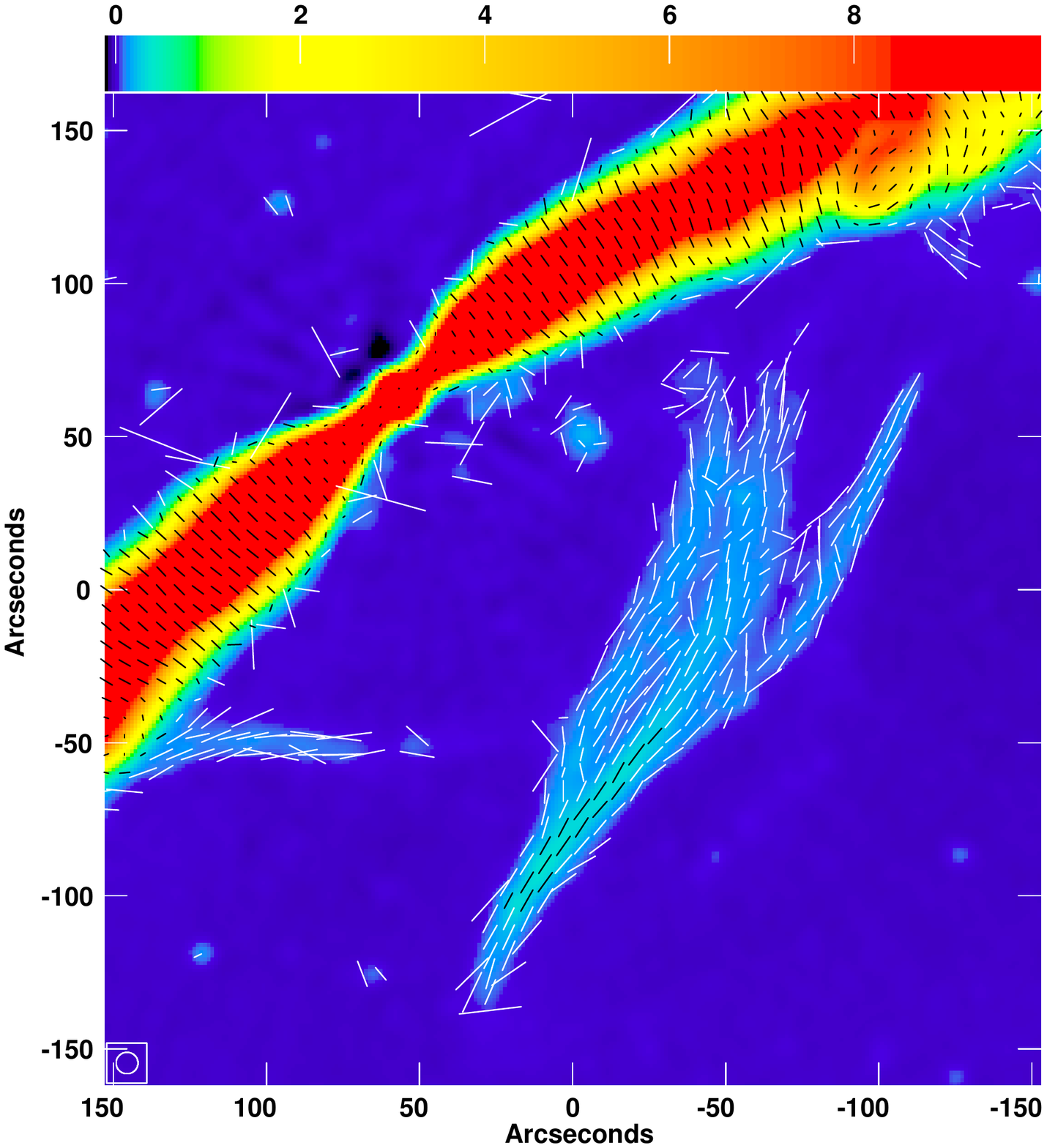}
}
\caption{Fractional polarization $\vec{B}$ vectors superposed on
  false-color total intensities for the core, jets, and threads. A
  vector length of $1\arcsec$ corresponds to 8\% fractional
  polarization, and total intensities in $\mathrm{mJy\,beam}^{-1}$ are
  indicated by the scale bar at the top. The FWHM restoring beam is
  shown by the white ellipse in the lower left corner.}
\label{fig:IC4296PolCore}
\end{figure}

\begin{figure}[!htb]
\centerline{
  \includegraphics[width=0.5\textwidth,trim = {1.cm 4.8cm 0.4cm 4.5cm},
    clip]{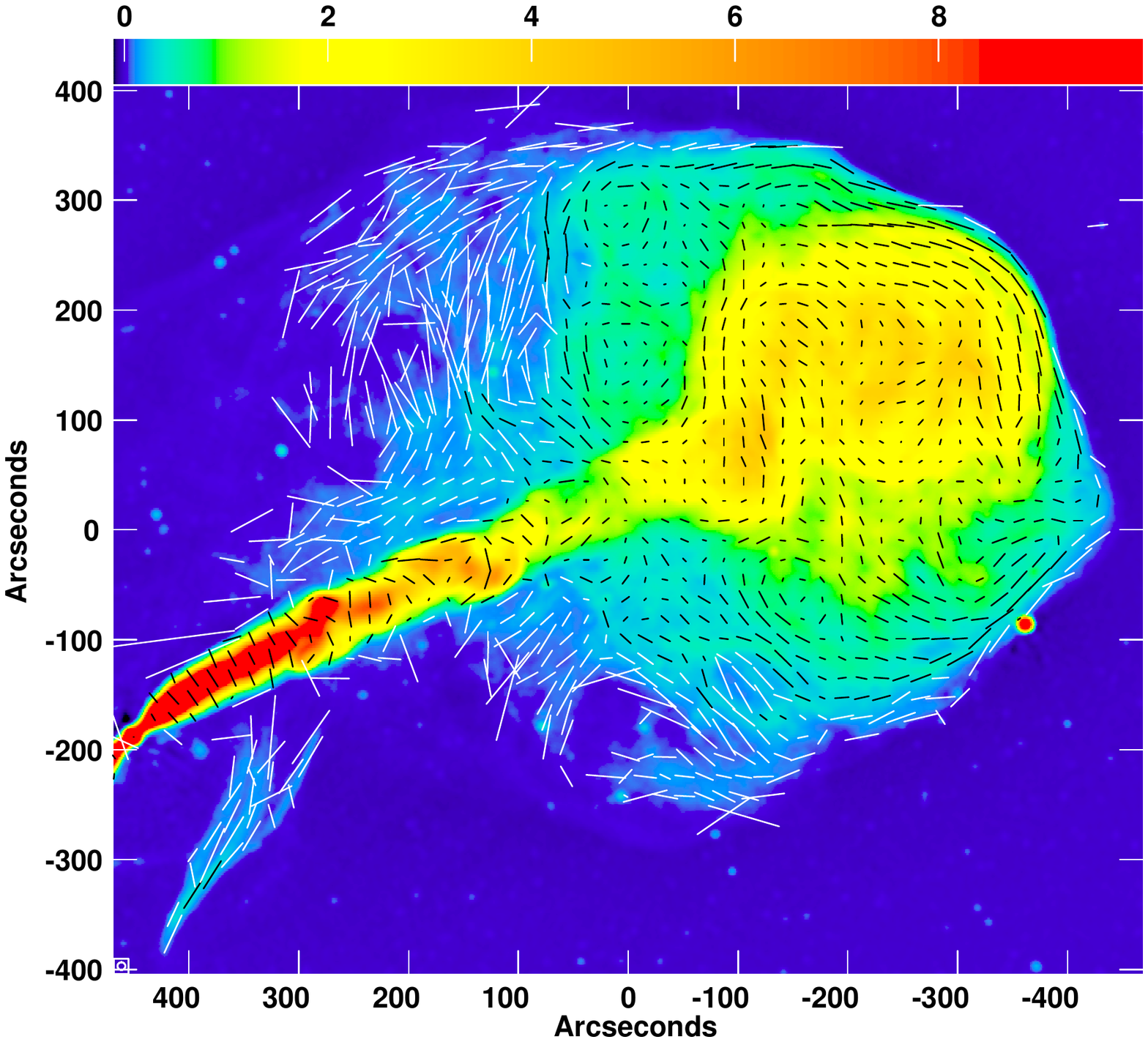}
}
\caption{Fractional polarization $\vec{B}$ vectors superposed on
  false-color total intensities for the northwest jet, lobe, and
  ribbon. A vector length of $1\arcsec$ corresponds to 2\% fractional
  polarization, and total intensities in $\mathrm{mJy\,beam}^{-1}$ are
  indicated by the scale bar at the top. The FWHM restoring beam is
  shown by the white ellipse in the lower left corner.}
\label{fig:IC4296NWLobe}
\end{figure}

\begin{figure}[!htb]
\centerline{
  \includegraphics[width=0.5\textwidth,trim = {1.cm 3.5cm 0.4cm 4.5cm},
    clip]{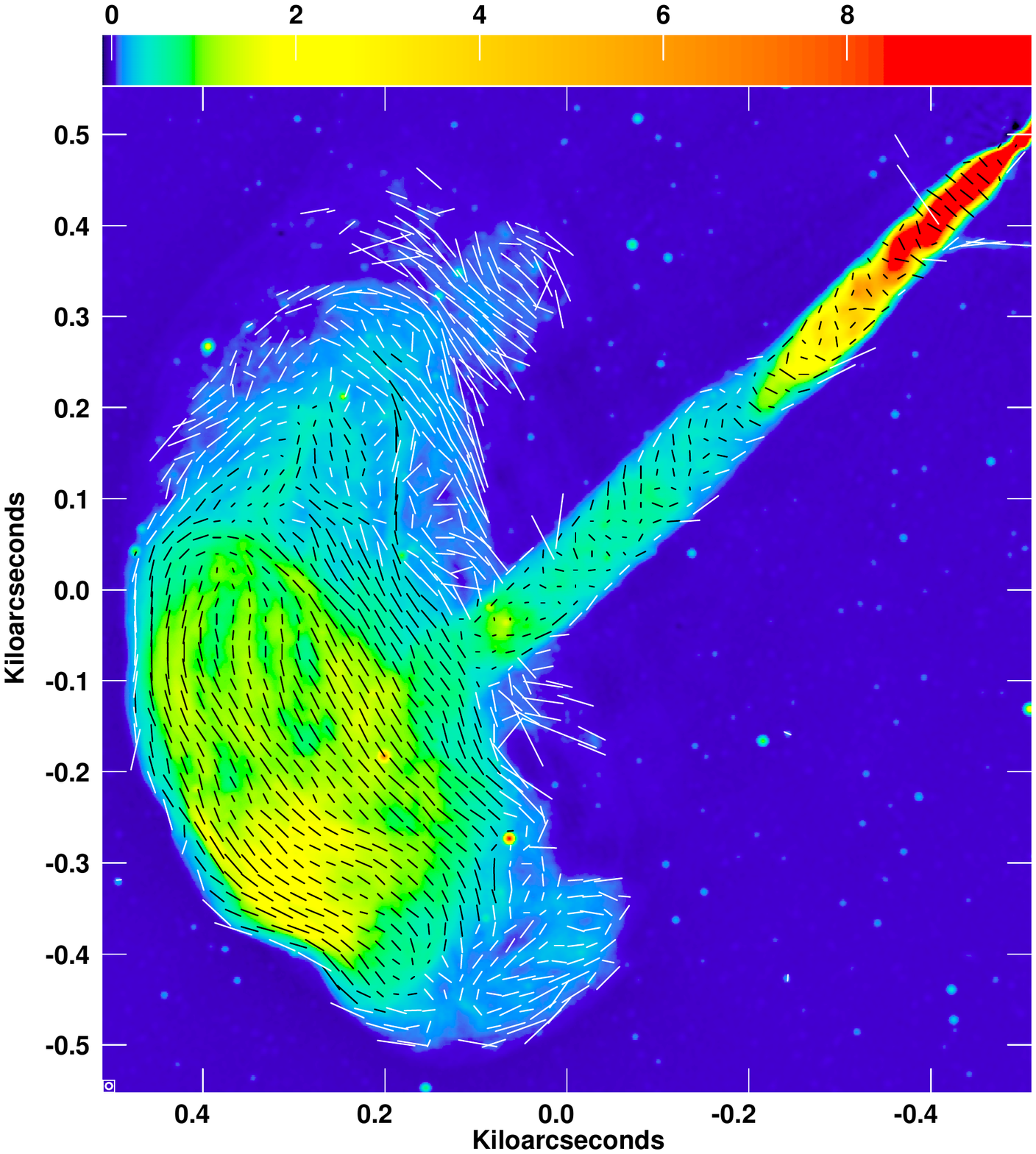}
}
\caption{Fractional polarization $\vec{B}$ vectors superposed on
  false-color total intensities for the southeast jet, lobe, and
  ribbon. A vector length of $1\arcsec$ corresponds to 2\% fractional
  polarization, and total intensities in $\mathrm{mJy\,beam}^{-1}$ are
  indicated by the scale bar at the top. The FWHM restoring beam is
  shown by the white ellipse in the lower left corner.}
\label{fig:IC4296SELobe}
\end{figure}

The polarized intensity image was corrected for positive noise bias
\citep[Appendix]{war74} and then divided by the total intensity to
yield fractional polarizations.  Stokes I false-color intensities and
fractional-polarization $\vec{B}$ vectors rotated to zero wavelength
tracing the direction of the magnetic field are shown in
Figures~\ref{fig:IC4296PolCore}--\ref{fig:IC4296SELobe}.  The vectors
are plotted as white or black lines for contrast only and do not have
different meanings.  The magnetic fields in the inner parts of the jets are
perpendicular to the jets, and velocity shear at the boundary layer
with static ambient gas appears to turn the $\vec{B}$ field parallel
to the jet boundaries.  Compression wraps $\vec{B}$ around the outer
edges of the lobes.  The fractional polarization in the jets ranges
from 10 to 30\%; in the brightest parts of the north-west lobe it is
typically $\gtrsim 10$\%, and 10 to 50\% in the south--east lobe.  The
magnetic fields run along the threads. The narrow threads are 30 to
70\% polarized, implying highly organized magnetic fields. Their
$\vec{B}$ fields are parallel to the threads, as expected from
velocity shear at the interface between the threads and the ambient
gas.  The ribbons between the jets and the lobes have lower
polarization and less-organized magnetic fields.

\hphantom{o}

Our MeerKAT images of IC~4296 are available in FITS format via the SARAO
archive \citep{images}.


\section{Radio Source Analysis}\label{sec:Discussion}

The 1.28\,GHz spectral luminosity of IC~4296 is
\begin{equation}
  L_\mathrm{1.28\,GHz} = \frac {4 \pi D_\mathrm{C}^2 S_\mathrm{1.28\,GHz}}
  {(1 + z)^{\alpha - 1}} \approx 4.5 \times 10^{24} \,\mathrm{W\,Hz}^{-1}~.
\end{equation}
For $\alpha = -0.7$, the 1.4\,GHz spectral luminosity is
$L_\mathrm{1.4\,GHz} \approx 4.3 \times 10^{24}\,\mathrm{W\,Hz}^{-1}$.
The spectral luminosity boundary separating FR\,\textsc{i} (center
brightened) from FR\,\textsc{ii} (edge brightened) radio sources
\citep{fan74} depends on the absolute $R$ magnitude of the host
galaxy, which is $R_{25} \mathrm{(Cousins)} = -23.3$ \citep{lau89}
for IC~4296.
The 1.4\,GHz luminosity of IC~4296 is $10 \times$ below the \citet{led96}
boundary, placing it in well inside the FR\,\textsc{i} region.
IC~4296 has always been classified as an FR\,\textsc{i} source in the
literature.  Our image in Figure~\ref{fig:IC4296IPolAll} shows
prominent FR\,\textsc{i} jets bright near the nucleus, but the
extended lobes are slightly edge-brightened.  The projected magnetic
field is predominantly transverse (i.e.~toroidal) in the inner jets, a
characteristic of FR\,\textsc{i} radio sources \citep{bri84}.

\subsection{The Core and Inner Jets}

The brightest point in our 1.28\,GHz MeerKAT image of IC~4296 is its
unresolved (deconvolved LAS $<3\arcsec$ FWHM) core whose peak flux
density is $S_\mathrm{p} = 188 \pm 7 \,\mathrm{mJy\,beam}^{-1}$.  Its
J2000 position $\alpha = 13^\mathrm{h}\,36^\mathrm{s}\,39\,\fs025 \pm
0\,\fs 011$, $\delta = -33^\circ\,57\arcmin\,57\,\farcs 01 \pm
0\,\farcs13$ matches both the ICRF position $\alpha = 13^\mathrm{h} \,
36^\mathrm{m} \, 39\,\fs 03275$, $\delta = -33^\circ \, 57\arcmin \,
57\,\farcs 0783$ \citep{fey15} and the 2MASX infrared position.  The
core flux density at 10\,GHz is $282 \pm 11$\,mJy \citep{ruf20}, and
the core spectral index $\alpha(1.28\,\mathrm{GHz},\,10\,\mathrm{GHz})
= +0.20$ implies significant synchrotron self-absorption.  For typical
kinetic temperatures $T \sim 10^{11}\,\mathrm{K}$ of synchrotron
electrons with critical frequencies near $\nu = 1.28\,\mathrm{GHz}$,
self-absorption implies a core brightness temperature approaching $T =
10^{11}\, \mathrm{K}$.  Thus the solid angle covered by the
flat-spectrum core of IC~4296 is not much more than $\Omega = 2 \times
10^{-6}\,\mathrm{arcsec}^2$ and is $\lesssim 1\,\mathrm{pc}$ in
extent. Also, any emission resolved by our MeerKAT image must be
very nearly transparent (optical depth $\tau \ll 1$) to synchrotron
self-absorption.

\citet{pel03} imaged the innermost jets of IC~4296 with
$5\,\mathrm{mas} \times 2\,\mathrm{mas}$ resolution at 8.4\,GHz and
detected the bases of both jets with a brightness ratio $R \sim 8$,
the northwest jet in PA $\approx -40^\circ$ being the
brighter.  For intrinsically similar jets and counterjets with bulk
speeds $\beta \equiv v/c$ and inclination angle $i < 90^\circ$ between
the line-of-sight and the approaching jet, Doppler boosting alone
produces a jet/counterjet brightness ratio \citep{bla79}
\begin{equation}\label{eqn:sideratio}
  R = \Biggl( \frac{1 + \beta \cos i} {1 -\beta \cos i} \Biggr)^{2-\alpha}~.
\end{equation}
For a jet spectral index $\alpha \geq -0.7$
(Figure~\ref{fig:IC4296SI}) and $\beta < 1$, the observed $R = 8$
suggests the parsec-scale approaching jet is inclined by $i <
69^\circ$.  Alternatively, we can drop the assumption that the jets
are intrinsically similar and use the flux-density ratio $R =
5.64\,\mathrm{Jy} / 3.73\,\mathrm{Jy} = 1.51$ of the
northwest/southeast lobes, which are presumably not Doppler boosted,
to estimate the intrinsic jet luminosities. Inserting $R = 8 / 1.51 =
5.3$ into Equation~\ref{eqn:sideratio} yields a more conservative
kinematic inclination limit $i < 73^\circ$.  In any case, the unique
value of these photometric estimates of relativistic jet orientation is to
determine  that the northwest jet is approaching.

On a slightly larger angular scale, \citet{dal09} used the \emph{HST}
to image the broad-line H$\alpha$ emission from IC~4296 = Abell
3565-BCG and found it to be spatially extended with FWHM $=
0\,\farcs13 \approx 30 \,\mathrm{pc}$. Their dynamical model matching
the velocity field of the broad-line gas yields the inclination angle
$i = 66\fdg0 ^{+3\fdg5}_{-3\fdg4}$ of the gas rotation axis. \citet{dal09}
also imaged the warped dust disk whose radius is $r \approx 1\farcs4
\approx 340 \,\mathrm{pc}$ and whose axial ratio indicates a
morphological inclination angle $i \approx 71^\circ$.  \citet{ruf20}
imaged the 230\,GHz CO(2--1) emission line in the central $r \approx
100\,\mathrm{pc}$ of the dust disk with ALMA, and they found the CO
disk rotation axis is inclined by $i = 68\,\fdg0 \pm 1\fdg5$.

If the inner jets are parallel to the rotation axes of the broad-line
gas and the dust disk, the sidedness ratio $R = 5.3$ yields estimates
of the jet velocity $\beta$ on mas scales ranging from $\beta = 0.6$
when $i = 60^\circ$ to $\beta = 0.9$ when $i = 71^\circ$.

Both radio jets brighten $\approx 4\arcsec \sim 1\,\mathrm{kpc}$ from
the nucleus \citep{ruf19a}, where their brightness ratio has fallen to
$R = 1.8 \pm 0.1$ \citep{ruf20}.  At this distance, most
FR\,\textsc{i} jets have already decelerated significantly
\citep{lai14}.  If the jet inclination angle is still $i \approx
68^\circ$ and $\alpha = -0.7$, then $R \leq 1.8$ suggests $\beta \leq
0.04$.  \citet{ruf20} assumed the values $\alpha = -0.6$ and $\beta =
0.75$ (typical of the innermost FR\,\textsc{i} jets) to estimate the
IC~4296 jet inclination angle $i = 81\,\fdg4 ^{+2\fdg7}_{-5\fdg2}$ on
kpc scales.  However, such a large inclination angle would imply that
jet bending projected along the line-of-sight is much greater than
bending projected onto the plane of the sky.
We consider that to be unlikely for jets that are so straight
projected onto the sky.  The JVLA 4.87\,GHz image
in \citet{ruf19a} fig.~1 shows that the jet and counterjet are
antiparallel to within $2^\circ$ up to $\approx 5\arcsec$ from the
nucleus.  At $r \sim 1\,\mathrm{kpc}$ the position angle of the
approaching jet is PA $= -50\,\fdg5 \pm 1\fdg0$ measured
east from north.

\subsection{The jets on 10\,kpc scales}

\begin{figure}[!htb]
\centerline{
  \includegraphics[width=0.5\textwidth,trim = {1.0cm 4.6cm 1.cm 4.4cm},
    clip]{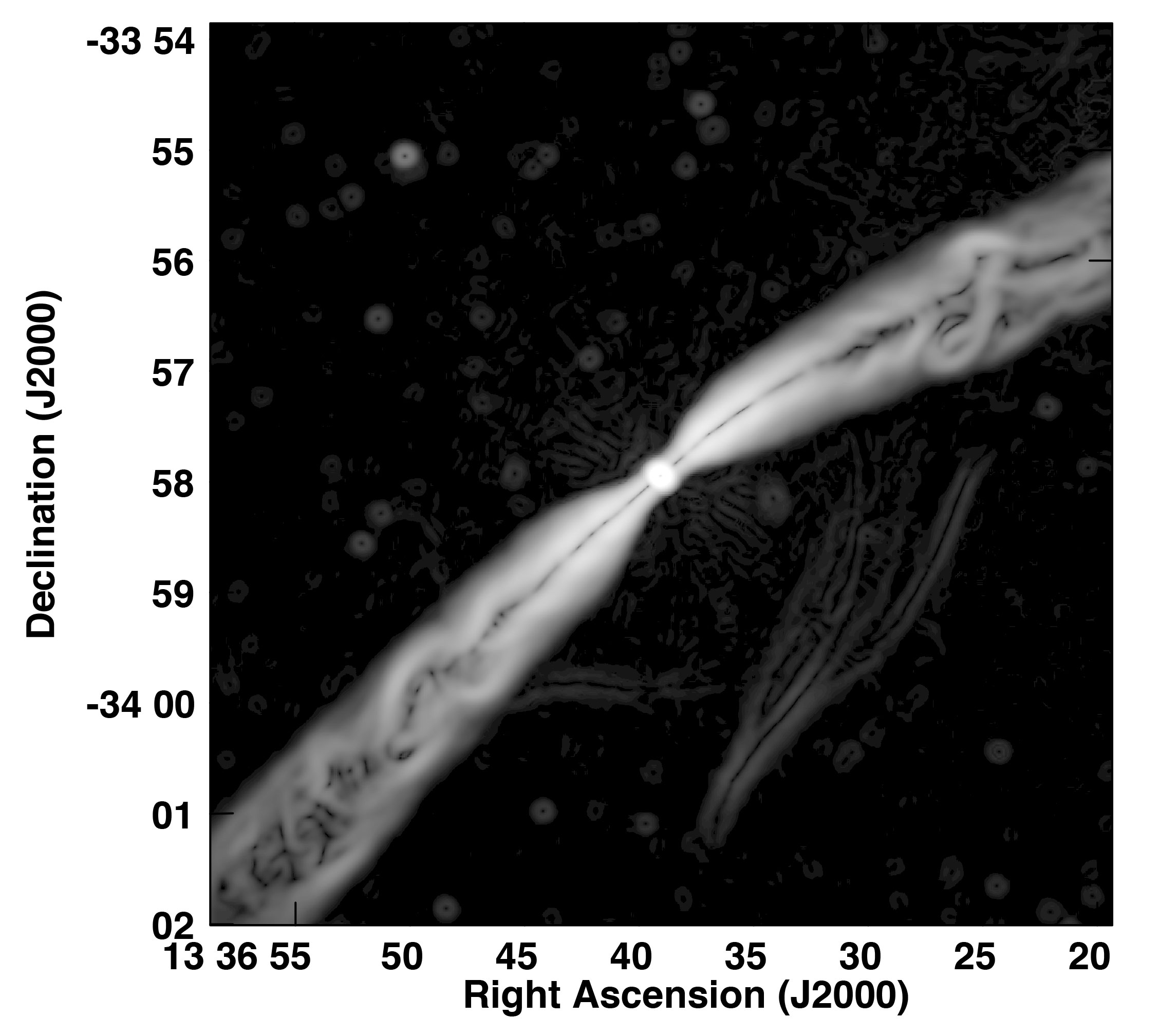}
}
\caption{Brightness in this gray-scale NINER image is proportional to
  the logarithm of the norm of the local intensity gradient, so jet
  ridge lines appear as dark lines, emission edges are lightened, and
  point sources look like donuts.  Three threads originate near the
  jet KH instability $\sim 2\farcm5 \approx 36\,\mathrm{kpc}$
  northwest of the bright radio core and one from the counterjet KH
  instability $\sim 2\farcm5$ southwest of the core.  The projected
  lengths of the threads are $\sim 50\,\mathrm{kpc}$.}
\label{fig:niner}
\end{figure}

To highlight structures within the jets resolved by our MeerKAT image,
we resampled the total-intensity image with small ($0\,\farcs24$)
pixels and used the \emph{AIPS} task NINER to apply a
3\,pixel$\,\times\,$3\,pixel Sobel filter whose output
(Figure~\ref{fig:niner}) is proportional to the norm of the local
intensity gradient.  Not all narrow radio features originating near
the core are jets, and \citet{bri84} ``ask that they contain a `spine'
of bright emission \dots before we call them \emph{jets}.''  In
Figure~\ref{fig:niner} the required spines appear as narrow dark
ridge lines in the center of the jets.  Out to $40\arcsec \approx
10$\,kpc from the core, the spines of both jets remain straight and
parallel to PA $= -49\,\fdg5 \pm 0\,\fdg5$. Their brightness ratios $R
\lesssim 1.3$ indicate little or no Doppler boosting.

\subsection{Threads}

Both jets begin to bend and then wiggle visibly $100\arcsec \approx
24\,\mathrm{kpc}$ from the core (Figures~\ref{fig:IC4296IPolAll},
\ref{fig:JetsThreads}, and \ref{fig:niner}).  Wiggles with wavelengths
several times the jet radius are probably caused by the helical ($n =
1$) normal mode of the Kelvin-Helmholtz (KH) instability driven by
velocity shear against the ambient medium \citep{har11} and appear
sinusoidal in projection onto the sky.  Brighter ``knots'' at the
turning points (e.g., at $\alpha =
13^\mathrm{h}\,36^\mathrm{m}\,25^\mathrm{s}$, $\delta =
-33^\circ\,56\arcmin$ in the northwest jet) about $2\,\farcm5$ from
the core may correspond to turning points where the helices are closer
to the line of sight and hence have a larger optical depth.
Long-wavelength KH jet oscillations promote mixing of the external
medium with the jet fluid and can catastrophically disrupt the jet
flow \citep{har11,har13}. Relativistic electrons escaping from the
jets may be the origin of the three faint narrow ``threads''
(Figure~\ref{fig:JetsThreads}, right panel) appearing
to originate from the KH knots in the northwest jet and the single
thread emerging from the southeast jet.

The threads must follow the ambient galaxy magnetic lines of force
because the gyro radius of the synchrotron electrons ($< 1
\,\mathrm{pc}$) is much smaller than the $\sim 1\,\mathrm{kpc}$ thread
radii and the magnetic fields are frozen into the ambient ionized
medium.  The threads originate around the half-light ``effective''
radius $r_\mathrm{e} = 37\,\mathrm{kpc}$ \citep{don11} of IC~4296,
where the rms density of $T = 10^7\,\mathrm{K}$ ($kT =
1\,\mathrm{keV}$) ambient electrons is $n_\mathrm{e} \approx 10^{-3}
\,\mathrm{cm}^{-3}$ \citep{gro19}.  The ambient pressure from such keV
protons and electrons is $P = 2 n_\mathrm{e} k T \approx 3 \times
10^{-12}\,\mathrm{dyne\,cm}^{-2}$.  The 0.5--5\,keV X-ray brightness
is highest inside  the red circle in the right panel of
Figure~\ref{fig:JetsThreads} \citep[matching the red
circle in][figure 4]{gro19}.
The higher thermal pressure inside the circle may explain why the
threads have avoided it.

The individual threads are marginally resolved with $\sim2$\,kpc
widths and have peak brightnesses $\sim
100\,\mu\mathrm{Jy\,beam}^{-1}$. Where the three nearly transparent
threads from the northwest jet merge (at least in projection), their
total brightness adds up to $\sim 300 \,\mu\mathrm{Jy\,beam}^{-1}$.
There may also be a very faint and marginally detected northern thread
barely visible in the right panel of Figure~\ref{fig:JetsThreads}
emerging at $\alpha = 13^\mathrm{h}\, 36^\mathrm{m}\,48^\mathrm{s}$,
$\delta = -33^\circ\,59\arcmin$.

As shown in Figures~\ref{fig:IC4296SI} and \ref{fig:IC4296PolCore},
the threads have $\langle \alpha \rangle \approx -1.2$ indicating
radiation losses and are highly polarized with longitudinal magnetic
fields.  The fractional polarizations are in excess of 50\% over the
bulk of the threads, indicating very ordered magnetic fields.  Clearly
these are coherent magnetic structures illuminated by an evolved
population of relativistic electrons.  The projected lengths of the
threads are 30--50\,kpc.

The minimum-energy  magnetic field strength
$B_\mathrm{min}$ of a synchrotron source can be estimated from its
brightness temperature $T_\mathrm{b}$, line-of-sight thickness $d$,
and proton/electron energy ratio $\kappa$ in the approximation that
the relativistic electrons emit at their critical frequencies from
$10^7$ to $10^{10}\,\mathrm{Hz}$ \citep{pac70}:
\begin{equation}\label{eqn:Bmin}
  \Biggl( \frac {B_\mathrm{min}} {\mu\mathrm{G}} \Biggr) \approx
  0.57 \Biggl[ \biggl( \frac {T_\mathrm{b}}{ \mathrm{K}} \biggr)
  \biggl( \frac {\nu (1+z)}{\mathrm{GHz}} \biggr)^{2-\alpha}
  \biggl( \frac {\mathrm{kpc}} {d} \biggr)
  (1 + \kappa)  \Biggr]^{2/7}
\end{equation}
The largest likely value of $\kappa$ is the proton/electron mass
ratio, so $\kappa \leq 2000$.  In marginally resolved threads with $d
\sim 2\,\mathrm{kpc}$, $S_\mathrm{p} = 0.1\,\mathrm{mJy\,beam}^{-1}$,
and $\alpha = -1.2$, $T_\mathrm{b} \approx 2\,\mathrm{K}$ and
$B_\mathrm{min} \lesssim 6\,\mu\mathrm{G}$.  The corresponding
particle pressure is $4/3$ times the $B^2/(8 \pi)$ magnetic pressure,
making the maximum total relativisitic pressure $P = 7 B^2 / (24 \pi)
\lesssim 3 \times 10^{-12} \,\mathrm{dyne\,cm}^{-2}$ comparable with
the ambient thermal pressure.  Thus the narrow threads are likely
confined and directed by the hot atmosphere of IC~4296.

\subsection{Ribbons}

The IC~4296 jets transition to ``ribbons'' approximately $400\arcsec =
100 \,\mathrm{kpc}$ from the core (Figure~\ref{fig:IC4296IPolAll}).
What we call ribbons differ from jets in that they are not
center-brightened and have no spines; instead they have nearly uniform
brightness within straight and narrow boundaries.  The southeast
ribbon extends to at least $700\arcsec \approx 180 \,\mathrm{kpc}$
from the core, where it becomes less distinct because it is seen in
projection against the southeast lobe.  It is less polarized than the
jet, and the magnetic field direction is not perpendicular to the
ribbon (Figure~\ref{fig:IC4296SELobe}).  Figure~\ref{fig:IC4296SI}
shows that the ribbon has a much steeper spectrum ($\alpha = -0.9$)
than the jet ($\alpha = -0.5$).  The northwest jet also transitions to
a ribbon, but the northwest ribbon is badly confused by emission from
the northwest lobe.

\begin{figure}[!htb]
\centerline{
  \includegraphics[width=0.5\textwidth,trim = {1.1cm 4.cm 1.cm 4.3cm},
    clip]{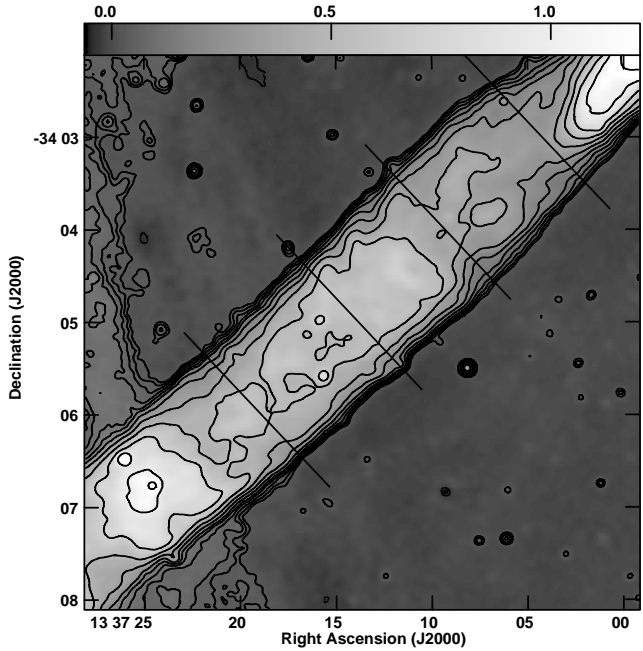}
}
\caption{This gray-scale image shows the southeast ``ribbon'' with a
  linear stretch from $-0.06$ to $+1.2\,\mathrm{mJy\,beam}^{-1}$. The
  contours are plotted at logarithmic intervals $T_\mathrm{b} =
  0.5\,\mathrm{K} \times \pm \,2^0,\,2^{1/2},\,2^1,\,2^{3/2},$\dots
  The four uniformly spaced diagonal lines mark four slices sampling
  brightness profiles normal to the ribbon.}
\label{fig:ribbon}
\end{figure}

\begin{figure}[!b]
\centerline{
  \includegraphics[width=0.5\textwidth,trim = {1.cm 9.5cm 6.2cm 8.cm},
    clip]{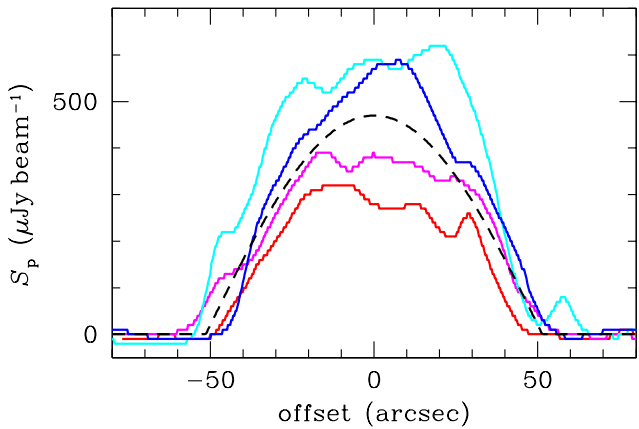}
}
\caption{The brightness profiles of the four slices marked in
  Figure~\ref{fig:ribbon} from left to right are shown by the blue,
  cyan, magenta, and red curves, respectively.  The dashed black curve
  is the cosine brightness profile of a nearly transparent
  circular tube with diameter $d = 25\,\mathrm{kpc}$ and constant
  volume emissivity.  }
\label{fig:slice}
\end{figure}

These properties of the southeast ribbon (Figure~\ref{fig:ribbon}) are
reminiscent of those of the radio cocoon surrounding the X-shaped
radio galaxy PKS~2014$-$55 \citep{cot20}.  They may be jet relics from
an an earlier time when the AGN was more active.  If the pressure in
the external medium is less than the internal pressure of the ribbon,
the ribbon will expand radially. The expansion speeds are typically
supersonic in the external medium but subsonic in the ribbons, giving
the ribbons time to reach internal pressure equilibrium \citep{beg84}
and become nearly circular tubes with nearly uniform volume
emissivity.  Their synchrotron optical depths are very low, so the
transverse brightness distribution of a circular tube with uniform
volume emissivity is $\cos(\rho)$, $-\pi/2 < \rho < +\pi/2$, where
$\rho$ is the angular offset from the center line of a tube with
radius $\rho = \pi /2$.

We extracted the brightness profiles from four uniformly spaced slices
(diagonal lines in Figure~\ref{fig:ribbon}) through the southeast
ribbon.  Figure~\ref{fig:slice} shows 
the observed profiles tend to be more flat-topped and edge-brightened
than the cosine profile, giving the appearance of thin, flat ribbons
rather than circular tubes.  The higher volume emissivity
near the tube walls suggests electron
re-acceleration by shocks at those walls.

Along the center line, $T_\mathrm{b} \approx 8\,\mathrm{K}$ and the
thickness of the circular tube is $d \approx 25\,\mathrm{kpc}$.  For
$\kappa < 2000$, Equation~\ref{eqn:Bmin} indicates a minimum-energy
magnetic field $B_\mathrm{min} \lesssim 4\,\mu\mathrm{G}$.  The
corresponding synchrotron radiative lifetime $\tau_\mathrm{syn} \sim
c_{12}B^{-3/2}$ \citep{pac70} is $\gtrsim 2 \times 10^8\,\mathrm{yr}$,
and the radiative lifetime for inverse-Compton (IC) scattering off the
$2.73(1+z)\,\mathrm{K}$ cosmic microwave background (CMB) is about twice
that.  Only synchrotron and IC-scattering radiation losses steepen the
radio spectra of the ribbons, so the large spectral-index difference
between the jets ($\alpha = -0.5$) and the ribbon ($\alpha = -0.9$) is
evidence that the ribbons are relics of jets that ceased activity $>
10^8\,\mathrm{yr}$ ago.

\begin{figure}[!htb]
\centerline{
  \includegraphics[width=0.5\textwidth,trim = {1.cm 3.cm 1.cm 3.cm},
    clip]{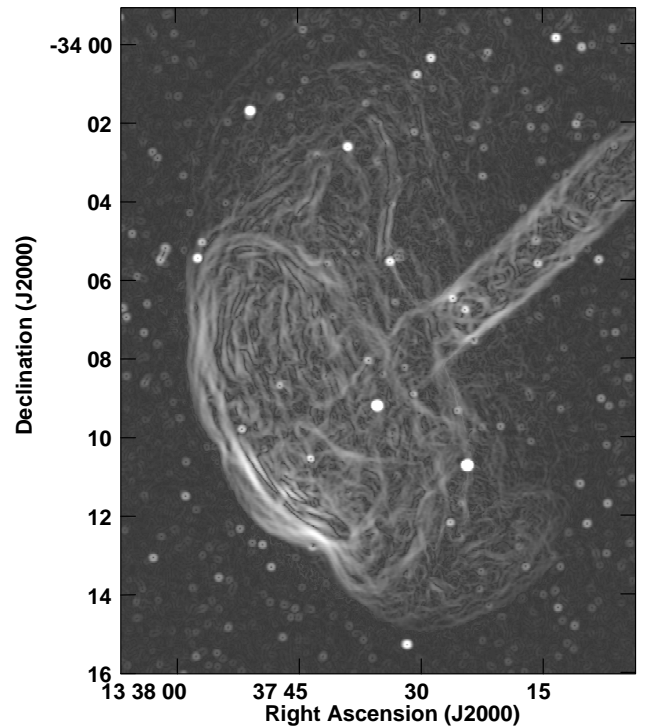}
}
\caption{This NINER image of the southeast lobe emphasizes the large
  ($9\arcmin \times 4\farcm5$, PA $= +28^\circ$) elliptical ring
  centered on $\alpha = 13^\mathrm{h}\,37^\mathrm{m}\,41^\mathrm{s}$,
  $\delta = -34^\circ \,09\arcmin$.}
\label{fig:ring}
\end{figure}


\subsection{Rings}

The southeast lobe and its backflow contains several elliptical
structures roughly centered on and perpendicular to the southeast jet
and ribbon as shown in Figure~\ref{fig:IC4296IPolAll} and in the NINER
Figure~\ref{fig:ring}.  The major axis of the brightest ellipse
extends from $\alpha = 13^\mathrm{h}\,37^\mathrm{m}\,31^\mathrm{s}$,
$\delta = -34^\circ 14\arcmin$ to $\alpha =
13^\mathrm{h}\,37^\mathrm{m}\,51^\mathrm{s}$, $\delta = -34^\circ
05\arcmin$.  If that ellipse is a circular vortex ring seen in
projection, its $2:1$ axis ratio indicates an inclination $i \approx
\arccos(0.5) = 60^\circ\pm 5^\circ$ or $120^\circ\pm 5^\circ$ from the
line-of-sight consistent with the expectation that the southeast
ribbon is a nearly straight extension of the southeast jet and lies
$\approx 30^\circ$ behind the plane of the sky.  The ring brightens the
outer edge of the southeast lobe and is further evidence that the
IC~4296 radio source was more luminous and had an FR\,\textsc{ii}
morphology in the distant past.

\section{Summary}\label{sec:Summary}

The unprecedented combination of angular resolution ($\theta \approx
7\arcsec$~FWHM), surface-brightness sensitivity ($\sigma \approx
5\,\mu\mathrm{Jy\,beam}^{-1} \approx 0.09 \,\mathrm{K}$ at $\nu =
1.28\,\mathrm{GHz}$), and high dynamic range in the new MeerKAT image of
IC~4296 revealed three new features: threads, ribbons,
and rings.  The threads are long ($\sim 50\,\mathrm{kpc}$), narrow
(width $\sim 2\,\mathrm{kpc}$), and faint ($\sim
0.1\,\mathrm{mJy\,beam}^{-1}$) synchrotron sources powered by
relativistic electrons escaping from helical Kelvin-Helmholtz
instabilities in the main radio jets.  The threads are highly
polarized with parallel magnetic fields, and they are likely confined
and directed by the higher-pressure ambient ionized gas and magnetic
fields of the host galaxy.  The ribbons begin where the main radio
jets end $\sim 100\, \mathrm{kpc}$ from the radio core.  Unlike
center-brightened FR\,\textsc{i} radio jets with well-organized
transverse magnetic fields, the ribbons have a nearly uniform
brightness distribution bounded by straight, sharp edges and
less-organized magnetic fields.  The IC~4296 ribbons are fainter
($\sim 0.5\,\mathrm{mJy\,beam}^{-1} \approx 8\,\mathrm{K}$ at
1.28\,GHz) than the jets, have steeper spectra ($\alpha \approx -0.9$
versus $\alpha \approx -0.5$) and appear to be relics of jet activity
that diminished $> 10^8\,\mathrm{yr}$ ago.  Near the leading edge of
the southeast radio lobe and roughly centered on the end of the
southeast ribbon is a thin vortex ring of emission observed projected
onto the sky as an ellipse with at $2:1$ axis ratio.  This axis ratio
is an orientation indicator showing that the jet/ribbon axis of
IC~4296 remains inclined by $\approx 60^\circ$ from the line-of-sight
all the way into the lobes.

Such morphological features are rare but may become common as
images with high angular resolution, brightness sensitivity, and
dynamic range are produced by the SKA precursor arrays MeerKAT and
the Australian SKA Pathfinder (ASKAP).  Recent examples include:
\begin{itemize}
\item{The collimated synchrotron threads linking the radio lobes of
  ESO 137-006 discovered in a MeerKAT image by \citet{ram20} are long
  and narrow like the IC~4296 threads, but their brightness can be
  much higher ($T_\mathrm{b} \sim 100\,\mathrm{K}$), their
  polarization has not been measured, and their origin is unknown.}
\item{Faint ($T_\mathrm{b} \approx 0.5\,\mathrm{K}$) cocoons
  surrounding the X-shaped radio galaxy PKS~2014$-$55 \citep{cot20}
  are backflow relics sharing many properties of the IC~4296 ribbons.
  These include smooth brightness distributions suggesting pressure
  equilibrium has been reached, and steep spectra indicating radiative
  aging of the synchrotron electrons.}
\item{Several Odd Radio Circles (ORCs) \citep{nor21} have
  been discovered in ASKAP's EMU pilot survey,
  although none are clearly vortex rings in radio lobes.}
\end{itemize}
\vfill\eject

\begin{acknowledgements}
The MeerKAT telescope is operated by the South African Radio Astronomy
Observatory, which is a facility of the National Research Foundation,
an agency of the Department of Science and Innovation.  The National
Radio Astronomy Observatory is a facility of the National Science
Foundation operated by Associated Universities, Inc.  SVW acknowledges
the financial assistance of the South African Radio Astronomy
Observatory (SARAO; \url{https://www.sarao.ac.za}). This research
has made use of the NASA/IPAC Extragalactic Database (NED), which is
funded by the National Aeronautics and Space Administration and
operated by the California Institute of Technology.  This research has
made use of the NASA/IPAC Infrared Science Archive, which is funded by
the National Aeronautics and Space Administration and operated by the
California Institute of Technology. This work has made use of data
from the European Space Agency (ESA) mission \emph{Gaia}
(\url{https://www.cosmos.esa.int/gaia}), processed by the \emph{Gaia}
Data Processing and Analysis Consortium (DPAC,
\url{https://www.cosmos.esa.int/web/gaia/dpac/consortium}). Funding
for the DPAC has been provided by national institutions, in particular
the institutions participating in the \emph{Gaia} Multilateral
Agreement.
\end{acknowledgements}

\facilities{Gaia, IRSA, MeerKAT}



\bibliographystyle{aasjournal}

\bibliography{arXiv1.bib}




\end{document}